\documentstyle{article}
\begin{document} 

\begin{flushright}
SUSSEX-TH-97-001\\
PU RCG-97/1\\
{\sf hep-th/9701082}\\
{\em Phys Rev D, in press}
\end{flushright}

\begin{center} 
\Large
{\bf Vacuum fluctuations in axion-dilaton cosmologies}\\ 
\bigskip
\large 
E.~J.~Copeland$^{1}$, Richard Easther$^{2,3,4}$ and  David Wands$^{5}$\\
\bigskip
\normalsize
$^{1}$Centre for Theoretical Physics, University of Sussex, Brighton,
BN1 9QH, U.K.\\ 
\vspace{.2cm} 
$^{2}$Newman Laboratory of Nuclear Studies, Cornell University, 
Ithaca NY14853, U.S.A. \\ 
\vspace{.2cm} 
$^{3}$Department of Mathematics, University of Waikato, Private Bag,
Hamilton, New Zealand \\ 
\vspace{.2cm} 
$^{4}$Department of Physics, Waseda University, 3-4-1
Okubo, Shinjuku-ku, Tokyo 169, Japan \\ 
\vspace{.2cm} $^{5}$School of Mathematical Studies, 
University of Portsmouth, Portsmouth PO1 2EG, U.K.
\end{center} 
\begin{abstract} 
We study axion-dilaton cosmologies derived from the low-energy string
effective action. We present the classical homogeneous
Friedmann-Robertson-Walker solutions and derive the semi-classical
perturbation spectra in the dilaton, axion and moduli
fields in the pre-Big Bang scenario. By constructing the unique
S-duality invariant field perturbations for the axion and dilaton fields
we derive S-duality invariant solutions, valid when the axion field is
time-dependent as well as in a  dilaton-vacuum cosmology. Whereas the
dilaton and moduli fields have steep blue perturbation spectra (with
spectral index $n=4$) we find that the axion spectrum depends upon the
expansion rate of the internal dimensions ($0.54 <n \leq 4$) which allows
scale-invariant ($n=1$) spectra. We note that for $n \leq 1$ the 
metric is non-singular in the conformal frame in which the axion is
minimally coupled.
\end{abstract}

\section{Introduction}

The dramatic progress that has been claimed in understanding black
holes in the context of string theory~\cite{BH,MALD} in the past year
focuses attention upon the implications that strings might have for
cosmology. The early universe provides a natural arena in which to
seek observational evidence for string theory as a fundamental
theory. General relativity describes all currently observed
gravitational physics with remarkable accuracy~\cite{Will93}, but its
description of the very early universe is expected to break down near
the Planck scale, and possibly at lower energies as well.

Most investigations of the cosmological consequences of string theory
have focussed on the role of the dilaton field, $\varphi$, which
provides a varying effective gravitational constant. In the absence of
other matter, the low energy action~\cite{effaction} leads to an
effective theory of gravity of the form proposed by Brans and
Dicke~\cite{BransDicke61}. Although the predicted value of the
Brans-Dicke parameter $\omega=-1$ is incompatible with present day
post-Newtonian tests~\cite{Will93}, it may be that loop corrections
yield an experimentally acceptable general relativistic
limit~\cite{Juan,DamourVilenkin95}, or that the dilaton's present day
value may be fixed by it acquiring a mass (for a detailed list of 
references see ~\cite{GSW}).

The pre-Big Bang scenario~\cite{pBB} is a specific example of a
cosmological model based on string theory which differs radically from
that predicted by general relativity because of the presence of the
dilaton field.  The weakly coupled, expanding, ``pre-Big Bang''
solutions have many similarities with conventional inflation
models~\cite{Deflation}, most notably the approach to flatness and
homogeneity on large scales by stretching the quantum vacuum state on
small scales up to large (super-horizon) scales. Solutions derived from
the low-energy effective action run into a ``Big Bang''
singularity~\cite{Bru+Ven94,Kaloper,EMW96} but the hope is that in the
full string theory, including higher-order terms, the pre-Big Bang
solution may be smoothly connected to a solution describing an
expanding, ``post-Big Bang'' universe~\cite{EM96,GV96} and, ultimately,
general relativistic evolution.

Our intention in this paper is to draw attention to the crucial role
that the antisymmetric tensor field,
$H_{abc}\equiv\partial_{[a}B_{bc]}$, may have in cosmological
scenarios based on the low-energy limit of the superstring action. An
antisymmetric tensor inevitably appears, along with the graviton and
dilaton, in the Neveu-Schwarz bosonic sector of the low-energy string
effective action,
\begin{equation}
\label{dilaction}
S = {1\over2\kappa_{\scriptscriptstyle{D}}^2}
 \int d^{\scriptscriptstyle{D}}x
 \sqrt{-g_{\scriptscriptstyle{D}}} \, e^{-\varphi}
 \left\{ R_{\scriptscriptstyle{D}} + (\nabla\varphi)^2 -
{1\over12} H^2
\right\} \, ,
\end{equation}
where $\kappa_{\scriptscriptstyle{D}}^2 \equiv 8\pi
G_{\scriptscriptstyle{D}}$, and $G_{\scriptscriptstyle{D}}$ is
Newton's constant in $D$-dimensions. Previous studies
\cite{Liddle}-\cite{Lu}
have considered the classical evolution of the antisymmetric tensor
field in simple cosmologies.  The role of the additional tensor fields
derived from the Ramond-Ramond sector of type IIA and type IIB string
theory has recently been considered in~\cite{Lukas,Lu}.  In this
paper we calculate for the first time the spectrum
of perturbations about the classical background field that may be
produced due to vacuum fluctuations in the antisymmetric tensor field.

We will consider spacetimes which contain a four dimensional
homogeneous and isotropic external metric. In four spacetime
dimensions the antisymmetric tensor field has only one degree of
freedom which may be represented by a pseudo-scalar axion field,
$\sigma$. For a homogeneous and isotropic metric we must have a
homogeneous axion field, $\sigma=\sigma(t)$.  This has been referred
to as the solitonic Anstaz due to the close connection with solitonic
p-brane solutions~\cite{Lukas}.  An alternative Ansatz, analogous to
elementary p-brane solutions~\cite{Lukas}, taking the tensor potential
$B_{\mu\nu}(t)$ to be time-dependent leads to an inhomogeneous axion
field and hence will be incompatible with a $D=4$
Friedmann-Robertson-Walker (FRW) cosmology~\cite{CLW95}.

In section~2 we review the classical evolution of FRW cosmologies both
with and without an evolving axion, stressing the important role of
the axion when the scale factor in the string frame becomes
small. Incorporating the evolution of the axion leads to a cosmology
that is very different from the classical pre-Big Bang scenario where
the axion field is fixed. We explore the differences between the
cosmological evolution as seen in different conformally related
metrics in section~3, drawing attention to the evolution in the axion
frame (where the axion field is minimally coupled). Specifically, we
find that the singular evolution of the metric in the string frame can
be non-singular in the axion frame.  These axion-dilaton solutions of
the low-energy action can be related by a duality transformation which
reduces to a scale factor duality in the absence of the axion field,
as described in section~4.

In section~5 we set out our formalism for describing inhomogeneous
linear perturbations about the homogeneous and isotropic 4-D background
solutions.  Even when the background axion field is set to zero, there
will inevitably be quantum fluctuations in the field. In section~6 we
calculate the spectrum of semi-classical axion perturbations as
well as dilaton and moduli perturbation spectra produced in the pre-Big
Bang scenario. In Section~7 we extend this calculation to more general
axion-dilaton cosmologies by constructing S-duality invariant
combinations of the field perturbations that enable us to derive
S-duality invariant solutions. Moreover we demonstrate that the
late-time dilaton and axion spectra turn out to be independent of the
preceding evolution along different but S-duality related classical
solutions. Importantly, the tilt of the axion spectrum can be
significantly different from the steep ``blue'' spectra of dilatons and
gravitons predicted by the pre-Big Bang scenario.

\section{Classical axion-dilaton cosmology}

Here we consider cosmological solutions derived from the
low-energy string action where we take the full $D$-dimensional
spacetime to have a metric of the form
\begin{equation}
ds_D^2 = -dt^2 + g_{ij}dx^idx^j
 + \gamma_{\scriptscriptstyle{IJ}}
    dX^{\scriptscriptstyle{I}}dX^{\scriptscriptstyle{J}} \, .
\end{equation}
$i$, $j$ run from $1$ to $3$, and $I$, $J$ run from $1$ to
$n=D-4$. We will allow for the variation of the $n$ compactified
dimensions by including a single modulus field, $\exp(n\beta)$,
proportional to the volume of this internal space, but neglect any
curvature or anisotropy so that $\gamma_{\scriptscriptstyle{IJ}} =
e^{2\beta}\delta_{\scriptscriptstyle{IJ}}$.

The effective dilaton in the four dimensional external spacetime is
then 
\begin{equation}
\phi \equiv \varphi- n \beta\, ,
\end{equation}
and the antisymmetric tensor field in four dimensions can be written    
in terms of the pseudo-scalar axion field $\sigma$ as
\begin{equation}
\label{axionansatz}
H^{abc} \equiv e^\phi \epsilon^{abcd} \nabla_d\sigma \, ,
\end{equation}
where $\epsilon^{abcd}$ is the covariant antisymmetric 4-form
such that $\nabla_e\epsilon^{abcd}=0$. 

The low-energy string effective action, given in Eq.~(\ref{dilaction}),
then becomes
\begin{eqnarray}
\label{4daction}
S & = & {1\over2\kappa^2} \int d^4x \sqrt{-g} e^{-\phi} \left\{ R +
 (\nabla\phi)^2
 - n (\nabla\beta)^2
 - {1\over2}e^{2\phi}(\nabla\sigma)^2 \right\} \, ,
\end{eqnarray}
where $\kappa^2\equiv 8\pi/M_{\rm Pl}^2$ determines the effective
value of the Planck mass when $\phi=0$, and $R$ is the Ricci scalar of
the four dimensional external spacetime.

We assume the external four dimensional spacetime is described by a
flat FRW metric with the line element 
\begin{equation}
\label{FRW}
ds^2 = a^2(\eta) \left\{ -d\eta^2 + \delta_{ij}dx^idx^j \right\}
 \, ,
\end{equation}
where $a(\eta)$ is the scale factor. In addition, FRW solutions
with non-zero spatial curvature can also be found~\cite{CLW94}.
To be compatible with a homogeneous and isotropic metric,
all the fields must be homogeneous and the action then
reduces to (up to a total derivative)
\begin{eqnarray}
S & = & {1\over2\kappa^2} \int d^3x \int d\eta e^{-\phi} \left\{
 -6a'^2 +6aa'\phi' - a^2\phi'^2
 + n a^2\beta'^2
 + {1\over2} e^{2\phi}a^2\sigma'^2 \right\} \, .
\end{eqnarray}

We refer to models with a constant axion field ($\sigma'=0$) as
{\em dilaton-vacuum\/} solutions. 
These are the well-known monotonic power-law
solutions\footnote {We shall not consider here the trivial flat
spacetime solution $\phi'=\beta'=a'=0$.}
\begin{eqnarray}
\label{dilphi}
e^\phi & = & e^{\phi_*} \left|{\eta\over\eta_*}\right|^{r_\pm}
 \, ,\\
\label{dila}
a & = & a_* \left|{\eta\over\eta_*}\right|^{(1+r_\pm)/2}
 \, ,\\
\label{dilbeta}
e^\beta & = & e^{\beta_*} \left|{\eta\over\eta_*}\right|^s \, ,
\end{eqnarray}
where the integration constants $r$ and $s$ determine the rate of
change of the effective dilaton and internal volume respectively. Note
that there is a constraint equation\footnote
{See Eq.~(\ref{h}) in the next section.}
which requires 
\begin{equation}
\label{r}
r_\pm = \pm \sqrt{3-2ns^2}\, .
\end{equation}
The dilaton-vacuum solutions are shown in Figs.~1 to~3.
If stable compactification has occurred and the volume of the
internal spaces is fixed ($s=0$, or $D=4$) we have
$r_\pm=\pm\sqrt{3}$. 

These dilaton-vacuum solutions can be expressed in terms of the proper time,
$t\equiv\int a\,d\eta$, giving
\begin{eqnarray}
\label{dilphit}
e^\phi & = & e^{\phi_*} \left|{t\over t_*}\right|^{2r_\pm/(3+r_\pm)}
 \, ,\\
\label{dilat}
a & = & a_* \left|{t\over t_*}\right|^{(1+r_\pm)/(3+r_\pm)}
 \, ,\\
\label{dilbetat}
e^\beta & = & e^{\beta_*} \left|{t\over t_*}\right|^{2s/(3+r_\pm)} \, ,
\end{eqnarray}
All these solutions have semi-infinite proper lifetimes. Those
starting from a singularity at $t=0$ for $t\geq0$, are denoted the
$(-)$ branch in Ref.~\cite{Bru+Ven94}, while those which approach a
singularity at $t=0$ for $t\leq0$ are referred to as the $(+)$
branch.

Our choice of origin for the time coordinate is arbitrary.
A more fundamental definition of the $(+/-)$ branches may be given by
considering the evolution of the shifted
dilaton~\cite{shifted,Tseytlin,EMW96}
\begin{equation}
\bar{\phi} \equiv \phi-3\ln(a) \, .
\end{equation}
Its time derivative,
\begin{equation}
\label{defbranch}
\bar\phi' = - \, {3+r_\pm \over 2\eta} \, ,
\end{equation}
is always positive on the $(+)$ branch (or $\eta<0$) due to the
constraint on the value of $r_\pm$ [Eq.~(\ref{r})], and always
negative on the $(-)$ branch (or $\eta>0$).  

These $(+/-)$ branches do {\em not} refer to the choice of sign for
$r_\pm$ in Eq.~(\ref{r}).  On either the $(+)$ or $(-)$ branches of
the dilaton-vacuum cosmologies we still have two distinct solutions
corresponding to the choice of the exponent $r_+$ or $r_-$, which
determines whether $\phi$ goes to negative or positive infinity,
respectively, as $\eta\to0$.

The generalisation of Eqs.(\ref{dilphi}--\ref{dilbeta}) to the more
general axion-dilaton cosmology where $\sigma'\neq0$ is
particularly simple~\cite{CLW94}:
\begin{eqnarray}
\label{axiphi}
e^\phi & = & {e^{\phi_*} \over 2} \left\{
\left|{\eta\over\eta_*}\right|^{-r} + \left|{\eta\over\eta_*}\right|^r
\right\} \, ,\\
\label{axia}
a^2 & = & {a_*^2\over2} \left\{ \left|{\eta\over\eta_*}\right|^{1-r}
 + \left|{\eta\over\eta_*}\right|^{1+r} \right\} \, ,\\
\label{axibeta}
e^\beta & = & e^{\beta_*} \left|{\eta\over\eta_*}\right|^s \, ,\\
\label{axisigma}
\sigma & = & \sigma_*
 \pm e^{-\phi_*} \left\{ { |\eta/\eta_*|^{-r} - |\eta/\eta_*|^r
\over |\eta/\eta_*|^{-r} + |\eta/\eta_*|^r } \right\} \, .
\end{eqnarray}
where the exponents are related via $r=\sqrt{3-2ns^2}$.
The time-dependent axion solutions are plotted in Figs.~1 to~4.
The presence of the axion places a lower bound on the value of the
dilaton, $\phi\geq\phi_*$. In doing so it interpolates between two
dilaton-vacuum solutions with an asymptotically constant axion field.
When $\eta\to\pm\infty$ the solutions approach the $r_+=+r$
dilaton-vacuum solution and as $\eta\to0$ the solution approaches the
$r_-=-r$ dilaton-vacuum solution. We shall see that the asymptotic
approach to dilaton-vacuum solutions at early and late times leads to a
particularly simple form for the semi-classical perturbation spectra,
independent of the intermediate evolution.

The dynamical effect of the axion field is negligible except near
$\eta\sim\eta_*$, when it leads to a bounce in the dilaton, $\phi'=0$.
Lukas {\em et al}~\cite{Lukas} have recently drawn attention to the
connection between these cosmological solutions and solitonic $p$-brane
solutions. If $r>1$ then this also leads to a bounce of the scale
factor, $a'=0$. However we still have the two disconnected branches,
as defined by Eq.~(\ref{defbranch}), corresponding to an increasing
shifted dilaton, approaching a singularity on the $(+)$ branch, or a
decreasing shifted dilaton, on the $(-)$ branch.

\section{Conformal frames}

Thus far we have written all the solutions in terms of the string
frame. If stringy matter is minimally coupled in this frame then
stringy test particles will follow geodesics with respect to this
metric. However, in order to understand the physical evolution
in these models it is revealing to look at conformally related
metrics, $g_{ab}\to\Omega^2g_{ab}$. If the conformal factor $\Omega^2$
is itself homogeneous then the transformed metric remains a FRW metric
but with scale factor $a\to\Omega a$.

This transformation of the scale factor can lead to some ambiguities
in the interpretation of the cosmological solutions~\cite{Deflation}.
For instance inflation is often defined as accelerated expansion
($\ddot{a}>0$). However such a definition is dependent upon the choice
of conformal frame in which one chooses to evaluate the acceleration.

Note that the proper time also changes under a conformal
transformation, $t\to\int\Omega dt$. One must not assume that a finite
proper time interval in one frame necessarily coincides with a finite
time in another frame and, in particular, we shall see that what looks
like a singular evolution in one frame may appear non-singular in
another frame.

\subsection{The Einstein frame}

By choosing a conformal factor $\Omega^2=e^{-\phi}$ we can work in a
frame\footnote
{All quantities calculated with respect to the Einstein metric will
carry a tilde.}
$\tilde{g}_{ab} = \Omega^2 g_{ab}$ where the dilaton is minimally
coupled to the external metric. 
Thus the gravitational part of the action in
Eq.~(\ref{4daction}) reduces to the Einstein-Hilbert
action~\cite{Deflation,Wands94}
\begin{equation}
S  =  {1\over2\kappa^2} \int d^4x \sqrt{-\widetilde{g}} \left\{
 \widetilde{R}
 - {1\over2} (\widetilde\nabla\phi)^2 
 - n (\widetilde\nabla\beta)^2
 - {1\over2} e^{2\phi}(\widetilde\nabla\sigma)^2 \right\} \, .
\end{equation}
and hence this is known as the Einstein frame\footnote
{Some authors refer to the frame where the full $D$-dimensional metric
is minimally coupled to the dilaton as the Einstein frame. This
corresponds to a Kaluza-Klein gravity theory, which will not in
general coincide with Einstein gravity in the four-dimensional
external metric.}.
Both the dilaton and the moduli fields have standard kinetic
terms in this frame and thus moduli particles and dilatons (in a
constant axion field) would therefore follow geodesics in the four
dimensional external spacetime of the Einstein frame.

The familiar field equations of general relativity
\begin{equation}
\widetilde{R}_{ab} - {1\over2}\widetilde{g}_{ab}\widetilde{R} =
\kappa^2 \widetilde{T}_{ab} \, ,
\end{equation}
apply in this frame. This may help one's cosmological intuition, which
is rooted in four dimensional general relativity, but also assists
mathematically by decoupling the equations for the evolution of the
metric from the value of $\phi$. The stress-energy tensor
\begin{equation}
\label{Tab}
\kappa^2\widetilde{T}_{ab}  =  {1\over2}
 \left(\widetilde{g}_a^c\widetilde{g}_b^d
 - {1\over2} \widetilde{g}_{ab}\widetilde{g}^{cd}\right) 
\ \left( \phi_{,c}\phi_{,d} + 2n\beta_{,c}\beta_{,d} +
 e^{2\phi}\sigma_{,c}\sigma_{,d} \right)
 \, .
\end{equation}
for homogeneous fields reduces to that for a perfect fluid with a
stiff equation of state~\cite{Tab+Taub,MW95}, i.e., pressure equal to
density,
\begin{equation}
\label{rho}
\widetilde{p}=\widetilde\rho = {1\over4\kappa^2} \left( \phi'^2
+2n\beta'^2 + e^{2\phi}\sigma'^2 \right) \, .
\end{equation}

The evolution equations for homogeneous fields in an FRW metric are then
\begin{eqnarray}
\label{phieom}
\phi'' + 2\widetilde{h}\phi' & = & e^{2\phi} \sigma'^2 \, , \\
\sigma'' + 2\widetilde{h}\sigma' & = & -2\phi'\sigma' \, , \\
\label{betaeom}
\beta'' + 2\widetilde{h}\beta' & = & 0 \, , \\
\label{hprime}
\widetilde{h}' & = & -{1\over6}
  \left( \phi'^2 + 2n \beta'^2 + e^{2\phi}\sigma'^2 \right) \, ,
\end{eqnarray}
plus the constraint
\begin{equation}
\label{h}
\widetilde{h}^2  =  {1\over12} 
  \left( \phi'^2 + 2n \beta'^2 + e^{2\phi}\sigma'^2 \right) \, ,
\end{equation}
where $\widetilde{h}\equiv\widetilde{a}'/\widetilde{a}$. 

{}From Eqs.~(\ref{hprime}) and~(\ref{h}) we see that 
in the general axion-dilaton case where we allow the dilaton,
axion and/or moduli fields to evolve, the expansion in the Einstein
frame always obeys $\widetilde{h}'+2\widetilde{h}^2=0$  
leading to the simple solution
for the scale factor
\begin{equation}
\label{Esf}
\widetilde{a} \equiv a e^{-\phi/2} = \widetilde{a}_* \left(
{\eta\over\eta_*} \right)^{1/2} \, .
\end{equation}
The scalar field equations of motion,
Eqs.~(\ref{phieom}--\ref{betaeom}), can then be integrated to give the
solutions presented in Eqs.~(\ref{axiphi}--\ref{axisigma}). 
Even in spatially curved FRW models the equations of motion remain
integrable~\cite{CLW94} despite the apparently non-trivial couplings
between the fields, because we can make a conformal transformation to
the Einstein frame where all the fields are minimally coupled to the
metric and, so long as they are all homogeneous, their combined
dynamical effect is no different from a single massless field~\cite{MW95}.

Because there is no interaction potential for the fields, the strong
energy condition ($\widetilde{p}>0$) is always satisfied [see
Eq.~(\ref{rho})] and the general relativistic singularity theorems
must hold. Thus we know there is no way to construct a non-singular
evolution in the Einstein frame with these massless fields.

In the string frame the usual general relativistic results do not hold.
Even without an interaction potential we can obtain an accelerated
expansion in Eq.~(\ref{dilat}) for the $(+)$ dilaton-vacuum branch
($t<0$) with $r_-<-1$ (or $r>1$ in the axion-dilaton solution). However
unlike conventional power-law inflation, $a\propto t^p$ with $p>1$, we
have ``pole-inflation'' with $p<0$, and we approach a curvature singularity
with $a\to\infty$ and $R\sim t^{-2}$ as $t\to0$.

In the Einstein frame we see that $\eta\to0$ on the $(+)$ branch always
corresponds to a collapsing universe with $\widetilde{a}\to0$. However
this still fulfils one definition of inflation, namely, that the
comoving Hubble length ($|d\widetilde{a}/d\widetilde{t}|^{-1} =
|\widetilde{a}/\widetilde{a}'|=2|\eta|$) decreases with
time~\cite{Deflation}. Thus a given comoving scale that starts
arbitrarily far within the Hubble scale in either conformal frame at
$\eta\to-\infty$ inevitably becomes larger than the Hubble scale in that
frame as $\eta\to0$. This allows one to produce perturbations in the
dilaton, moduli and graviton fields on scales much larger than the
present Hubble scale from quantum fluctuations in flat spacetime at
earlier times, as we shall discuss in more detail later.

In both the string frame and the Einstein frame we either reach a
curvature singularity in a finite proper time in the future for $\eta<0$
or emerge from a curvature singularity at a finite time in the past for
$\eta>0$. The only exception to this is the axion-dilaton solutions in
the string frame when $r=1$ ($s=\pm\sqrt{1/n}$) in
Eqs.~(\ref{axiphi}--\ref{axisigma}) which bounce at exactly
$\eta=0$~\cite{Behrndt}. However even in this case the dilaton and
moduli become infinite at a finite proper time.

\subsection{The axion frame}

Both the dilaton and moduli fields are minimally coupled in the
Einstein frame (i.e. they have standard kinetic terms). However the
axion's kinetic term retains a non-minimal coupling to the
dilaton. This can be removed by a conformal transformation to another
conformally related metric, the ``axion frame'', given by
$\bar{g}_{ab}=e^{-2\phi}g_{ab}$ and hence
\begin{equation}
\label{bara}
\bar{a} \equiv e^{\phi/2} a\equiv e^\phi\widetilde{a} \, .
\end{equation}
The axion field is a minimally coupled massless scalar field in this
frame and thus axionic test particles would follow geodesics with
respect to this metric.
Although conformally related to the string and Einstein frames, the
metric the axions ``see'' may behave very differently from the metrics 
in the string or
Einstein frames.

In terms of conformal time, the axionic scale factor for the
dilaton-vacuum solutions given by Eqs.~(\ref{dilphi}) and~(\ref{dila})
when $\sigma'=0$, evolves as
\begin{equation} 
\label{axionsf}
\bar{a} = \bar{a}_* \left( {\eta\over\eta_*} \right)^{r_\pm+(1/2)} \, . 
\end{equation} 
We see that the proper time in the axion frame is given by 
\begin{equation}
\bar{t}\equiv\int\bar{a}\,d\eta \sim |\eta|^{r_\pm+(3/2)} \, , 
\end{equation} 
so it takes an infinite proper time to reach $\eta=0$ for $r_-<-3/2$
(and thus $ns^2<3/8$) and the scalar curvature for the axion metric,
$\bar{R}\sim\bar{t}^{-2}$, vanishes as $\eta\to0$. However, these
same dilaton-vacuum solutions then reach $\eta\to\pm\infty$ in a finite
proper time where $\bar{R}$ diverges. Because the conformal factor
diverges as $\eta\to0$ it stretches out the curvature singularity in the
string metric into a non-singular evolution in the axion frame. But as
$\Omega^2=e^\phi\to0$ as $\eta\to\pm\infty$ the non-singular evolution in
the string frame gets compressed into a curvature singularity in the
axion frame.

Similar behaviour has previously been noted in the case of black holes
in the low-energy limit of string theory~\cite{Sdual}. 
Astronauts made of axionic matter falling into an axion-dilaton black
hole in $D=4$ would take an infinite proper time (measured by their
axionic clocks) to fall into what appears, in the Einstein frame, to
be the singularity.

In terms of the proper time $\bar{t}$ in the axion frame we have
\begin{equation}
\bar{a} = \bar{a}_* \left( {\bar{t}\over\bar{t}_*}
 \right)^{(1+2r_\pm)/(3+2r_\pm)}
\end{equation}
For $r_-<-3/2$ we have conventional power-law inflation (not
pole-inflation) with $\bar{a}\sim\bar{t}^{\bar{p}}$ with
$\bar{p}=1+[2/(-2r_--3)]>1$. We shall see that this has important
consequences for the tilt of the power-spectrum of semi-classical
perturbations in the axion field produced on large scales.

These dilaton-vacuum solutions still have a curvature singularity as
$\bar{t}\to0$ so the solutions still have only a semi-infinite
lifetime in the axion frame\footnote{The case $r_-=-3/2$ is an
exception as they correspond to de Sitter expansion
with $\bar{a}\sim \exp(H\bar{t})$.}, but for $r_-<-3/2$ this now
coincides with $\eta\to\pm\infty$, so the identification of the $(+)$
and $(-)$ branches as solutions approaching or leaving a singularity
is interchanged for the $r_-$ solution in the axion frame when
$r_-<-3/2$. 

However this implies that the axion-dilaton solutions with a
time-dependent axion field will be non-singular in the axion frame if
$r>3/2$. Remember that the axion-dilaton solutions 
with $\sigma'\neq0$ match $r_-=-r$ dilaton-vacuum solutions at
$\eta\to0$ onto $r_+=+r$ solutions as $|\eta|\to\infty$. Thus for
$r>3/2$, the axion-dilaton solutions are non-singular in the axion
frame as $\eta\to0$ (because $-r<-3/2$) {\em and} non-singular as
$\eta\to\infty$ (because $+r>-3/2$).

The general evolution for the axion-dilaton system,
Eqs.~(\ref{axiphi}--\ref{axisigma}) is given in the axion frame by
Eq.~(\ref{bara}), so
\begin{equation}
\label{generalaxionsf}
\bar{a} = {\bar{a}_*\over2}
 \left\{ \left|{\eta\over\eta_*}\right|^{(1/2)-r}
 + \left|{\eta\over\eta_*}\right|^{(1/2)+r} \right\} \, .
\end{equation}
{}From this we can extract the proper time, $\bar{t}$,
\begin{equation}
\label{genaxiontime}
{\bar{t} \over \bar{t}_*} = \left( {9-4r^2 \over 12} \right)
\left[ {2 \over 3-2r}\left|{\eta\over\eta_*}\right|^{(3/2)-r}
 + {2 \over 3+2r}\left|{\eta\over\eta_*}\right|^{(3/2)+r}
\right]
\end{equation}
which, as shown above, is semi-infinite
($0\leq\bar{t}/\bar{t}_*<\infty$) for $r<3/2$ but unbounded
($-\infty<\bar{t}/\bar{t}_*<\infty$) for $r\geq 3/2$. 
Equations~(\ref{generalaxionsf}) and~(\ref{genaxiontime}) give us a
parametric solution for the axion frame scale factor in terms of the
proper time in that frame. 

Representative examples showing the behaviour of $\bar{a}(\bar{t})$ for
different values of $r$ are given in Fig.~6. Note that the scale factor,
$\bar{a}$, has a non-zero minimum value (i.e., a bounce) whenever
$r>1/2$. When $r>3/2$, $\bar{a}$ becomes infinite as $\bar{t}
\rightarrow -\infty$, passes through a non-zero minimum value and then
expands indefinitely as $\bar{t}\rightarrow \infty$. In particular, if
stable compactification has occured so that the moduli field is fixed
($s=0$), or if $D=4$ (so that $n=0$), then $r=\sqrt{3}$ and the
axion-dilaton solution is {\em always} non-singular in the axion frame.
When $1/2<r<3/2$, $\bar{a}$ does have a bounce but is singular, since it
becomes infinitely large in a finite proper time, as $\bar{t}\to0$.
Finally, when $r<1/2$, the solutions are monotonic and there is a
singularity when $\bar{a}$ vanishes at $\bar{t}=0$.

\section{Duality}

\subsection{Scale-factor duality}

The constant axion solutions given in
Eqs.~(\ref{dilphit})--(\ref{dilbetat}) are related by the scale factor
duality transformation~\cite{shifted,sfduality}
\begin{equation}
\label{SFdual}
a\to {1\over a} \, , \qquad e^\phi \to {e^\phi\over a^6}
\end{equation}
which corresponds to a change in the parameters
\begin{equation}
\label{rSFdual}
a_* \to {1\over a_*}\, , \quad e^{\phi_*} \to {e^{\phi_*}\over a_*^6} \,, 
\quad r_\pm \to -\, {3+2r_\pm \over 2+r_\pm} \, ,
\end{equation}
in Eqs.~(\ref{dilphit}--\ref{dilbetat}).
This is a particular case of a more general $O(d,d)$
duality~\cite{sfduality} where the axion field remains constant.  When
we can neglect the evolution of the moduli fields ($s=0$ and hence
$r_\pm=\pm\sqrt{3}$) this coincides with $r_\pm\to r_\mp$. Note that
this scale factor duality does {\em not} take one from the $(+)$ to
$(-)$ branch or vice versa. This would require a time reversal.

The pre-Big Bang scenario postulates a non-singular universe by linking
the semi-infinite lifetime expanding $(+)$ branch, $a\sim(-\eta)^{-p}$,
starting at $\eta=-\infty$ to the semi-infinite expanding $(-)$ branch,
$a\sim\eta^p$, travelling off to $\eta=+\infty$ via a scale factor duality
transformation, plus time reversal near, the singularity at
$\eta\approx0$. 

The presence of a time-dependent axion field $\sigma(t)$ (the
solitonic Ansatz) breaks the $O(d,d)$-invariance which requires that it
is the antisymmetric potential $B_{ab}$ which is homogeneous (the
elementary Ansatz). The elementary Ansatz  is only compatible for a
restricted class of metrics in anisotropic 4-D
spacetimes~\cite{CLW95,BarrowET1996}, or if the axion is constant.

\subsection{S-duality}

Solutions with a time-dependent axion field do respect the
invariance of the low energy string action under the $SL(2,R)$
transformation:
\begin{equation}
\label{Sduality}
\lambda \to {\alpha\lambda+\beta \over \gamma\lambda+\delta}
\, ,
\end{equation}
where $\alpha$, $\beta$, $\gamma$ and $\delta$ are real constants
subject to $\alpha\delta-\beta\gamma=1$,
and $\lambda$ is the complex dilaton field
\begin{equation}
\lambda = \sigma + ie^{-\phi} \, .
\end{equation}
This leads to
\begin{eqnarray}
\label{Saxiphi}
e^\phi &\to& \gamma^2 e^{-\phi} + (\delta+\gamma\sigma)^2e^\phi \,,\\
\label{Saxisig}
e^\phi\sigma &\to& (\beta+\alpha\sigma)(\delta+\gamma\sigma)e^\phi
 + \alpha\gamma e^{-\phi} \,.
\end{eqnarray}

In the underlying string theory this represents the modular
invariance of the complex dilaton~\cite{GSW}, but working only with the
classical fields in the low energy action, it represents a
transformation between the dilaton and axion fields which leaves
invariant
\begin{equation}
dS^2 \equiv e^{2\phi} d\lambda d\lambda^* =
 d\phi^2+e^{2\phi}d\sigma^2\,.
\end{equation}

In terms of $\phi$ and $\sigma$, or indeed $\lambda$, it is 
not immediately apparent that $dS^2$ should remain 
invariant under 
the transformation given in Eq.~(\ref{Sduality}). 
It is rather more transparent if we define the matrix
\begin{equation}
\label{N}
N = \left(
\begin{array}{cc}
e^\phi & e^\phi \sigma \\
e^\phi \sigma & e^{-\phi} + e^\phi \sigma^2
\end{array}
\right) \,,
\end{equation}
which obeys $N^TJN=J$, where
\begin{equation}
J = \left(
\begin{array}{cc}
0 & 1 \\
-1 & 0
\end{array}
\right)
\,,
\end{equation}
and thus is a member of $SL(2,R)$. The particular $SL(2,R)$ 
transformation given in Eq.~(\ref{Sduality}) is given by
\begin{equation}
N \to \Theta N \Theta^T \,,
\end{equation}
where
\begin{equation}
\Theta = \left(
\begin{array}{cc}
\delta & \gamma \\
\beta & \alpha
\end{array}
\right) \,,
\end{equation}
is also a member of $SL(2,R)$.  Then we can write $dS^2=$tr$(JdNJdN)/2$
and,  noting that $\Theta^TJ\Theta=J$, it  is straightforward to verify
that $dS^2$ is invariant.  We will also find this notation particularly
convenient  later to construct explicitly S-duality invariant dilaton
and axion field perturbations.

The Lagrange density of the axion and dilaton fields in the
Einstein frame
\begin{equation}
-{1\over4}{\rm tr}(J\nabla_\mu N J\nabla^\mu N) =
 - {1\over2} (\widetilde\nabla\phi)^2
 - {1\over2} e^{2\phi}(\widetilde\nabla\sigma)^2 
 \,.
\end{equation}
is S-duality invariant and, because the dilaton and axion are 
minimally coupled to the other fields
in the Einstein frame, the evolution of the moduli field, $\beta$, and
scale factor, $\widetilde{a}$, are unaffected by the S-duality
transformation.  However the scale factor in the original string frame
must transform and will not remain invariant under a non-trivial  
transformation. 

If we choose $\gamma/\delta=-1/\sigma_*$, when
$\sigma=\sigma_*=$constant, the solutions given in
Eqs.~(\ref{dilphi})--(\ref{dilbeta}) are mapped by the transformation
in Eq.~(\ref{Sduality}) to
\begin{eqnarray}
e^\phi &\to& \gamma^2 e^{-\phi} \, ,\\
\sigma &\to& {\alpha\over\gamma}\, , \\
a &\to& \gamma e^{-\phi} a  \, ,
\end{eqnarray}
which leaves $\sigma$ constant. In particular, for $\gamma^2=1$ we have
$\phi\to-\phi$ and hence this is a transformation between strong and
weak coupling. The form of the solutions given in
Eqs.~(\ref{dilphi})--(\ref{dilbeta}) are unchanged, but the parameters
\begin{equation}
e^{\phi_*} \to e^{-\phi_*} \,, \quad a_* \to e^{-\phi_*}a_* \,,
 \quad r_\pm \to r_\mp \, .
\end{equation}
Comparing with Eq.~(\ref{rSFdual}), we find that in the particular
case when $ns^2=0$, and hence $r_\pm=\pm\sqrt{3}$, this coincides with the
scale factor duality given in Eq.~(\ref{SFdual}).

The more general S-duality transformations of $e^\phi$ and $\sigma$
given in Eqs.~(\ref{Saxiphi}) and~(\ref{Saxisig}) can be shown to
relate the dilaton-vacuum cosmologies, given in
Eqs.~(\ref{dilphi}--\ref{dilbeta}), to the more general axion-dilaton
cosmologies with a time dependent axion field, given in
Eqs.~(\ref{axiphi}--\ref{axisigma}), with a fixed value of
$r=|r_\pm|$. Thus the S-duality transformation allows one to generate
the general axion-dilaton solutions with a given value of $r$ starting
from only with the dilaton-vacuum solution with $r_\pm=\pm r$.

\section{Linear perturbations}

Thus far we have considered only homogeneous classical solutions to the
equations of motion. In the next section we will consider inhomogeneous
perturbations that may be generated due to vacuum fluctuations. In order
to follow their evolution we will set up in this section the formalism
required to describe linear perturbations about the homogeneous
background metric. 

We shall consider perturbations of the four-dimensional metric in the
spatially flat gauge\footnote{Called the ``off-diagonal gauge'' in
Ref.~\cite{BrusteinET1995}.} (or in more general FRW models, the uniform
spatial curvature gauge~\cite{Hwang}), using the Einstein frame, so that
to first-order the perturbed line element can be written as
\begin{equation}
\label{dds}
d\widetilde{s}^2  =  \widetilde{a}^2(\eta)
 \left\{ -(1+2\widetilde{A})d\eta^2
 + 2\widetilde{B}_{,i} d\eta dx^i
+ \left[\delta_{ij} + h_{ij}\right] dx^i dx^j \right\} \, ,
\end{equation}
where $\widetilde{A}$ and $\widetilde{B}$ are the scalar metric
perturbations (in the notation of Ref.~\cite{Bardeen80}) and $h_{ij}$
represents a transverse and traceless tensor perturbation.  
Linear perturbations about the homogeneous background fields can be
decomposed as a sum of Fourier modes with comoving wavenumber $k$ (and
in the case of the tensor perturbations two independent
polarisations) which evolve independently of other wavenumbers.

\subsection{Scalar metric perturbations}

The advantage of splitting the metric perturbations into scalar and
tensor parts is that the scalar and tensor modes evolve independently
to first order with only the scalar perturbations being coupled to
scalar field fluctuations~\cite{Bardeen80}. In the spatially flat
gauge we have the added simplification that the evolution equations
for linear perturbations about homogeneous scalar fields are decoupled
from the metric perturbations, although they are still related by a
constraint equation.
 
The field equations for the linearised scalar perturbations are
\begin{eqnarray}
\label{dphieom}
\delta\phi'' + 2\widetilde{h}\delta\phi' + k^2\delta\phi
 & = & 
2e^{2\phi}\sigma'^2\delta\phi + 2 e^{2\phi}\sigma'\delta\sigma' \\
\label{dsigmaeom}
\delta\sigma''+2\widetilde{h}\delta\sigma' + k^2\delta\sigma
 & = & 
- 2(\sigma'\delta\phi'+\phi'\delta\sigma') \\
\label{dbetaeom}
\delta\beta'' +2\widetilde{h}\delta\beta' +k^2\delta\beta
 & = & 0 \, ,\\
\label{Aeom}
\widetilde{A}'' + 2\widetilde{h}\widetilde{A}' + k^2 \widetilde{A}
 & = & 0 \, .
\end{eqnarray}
plus the constraints
\begin{eqnarray}
\label{AofB}
\widetilde{A} &=& - ( \widetilde{B}' +2\widetilde{h}\widetilde{B} ) \, ,\\
\label{Aconstraint}
 &=& {\phi'\over4\widetilde{h}} \, \delta\phi
 +{e^{2\phi}\sigma'\over4\widetilde{h}} \, \delta\sigma
  +{n\beta'\over2\widetilde{h}} \, \delta\beta \, .
\end{eqnarray}

Note that the scalar metric perturbations are not invariant under a
conformal transformation. Even the spatially flat nature of the 
line element in Eq.~(\ref{dds}) is not preserved under a conformal
transformation back to the string frame due to the first-order
perturbation in the conformal factor $e^\phi=e^{\phi_0}(1+\delta\phi)$.
However the tensor perturbation remains invariant under both conformal
transformations and gauge transformations $\eta\to\eta+\delta\eta$. 

The evolution equation for the scalar metric perturbations,
Eq.~(\ref{Aeom}) is independent of the evolution of the different
scalar fields and is dependent only on the evolution of the Einstein
frame scale factor $\widetilde{a}(\eta)$ given by
Eq.~(\ref{Esf}). This in turn is determined solely by the stiff fluid
equation of state for the homogeneous fields in the Einstein frame,
regardless of the time dependence of the axion field.
Equation~(\ref{Aeom}) can be integrated to give the
general solution
\begin{equation}
\label{genA}
\widetilde{A} = 
 \left[ A_+ H_0^{(1)}(-k\eta) + A_- H_0^{(2)}(-k\eta) \right] \, ,
\end{equation}
where $H_\nu^{(1)}(z)\equiv J_\nu(z)+iY_\nu(z)$ and $H_\nu^{(2)}\equiv
J_\nu(z)-iY_\nu(z)$ are  Hankel functions of the 
first and second kind. Using the recurrence relation between Bessel
functions, we obtain from Eqs.~(\ref{AofB}) and~(\ref{genA}),
\begin{equation}
\label{genB}
\widetilde{B} = {1\over k}
 \left[ A_+ H_1^{(1)}(-k\eta) + A_- H_1^{(2)}(-k\eta) \right] \, .
\end{equation}

Our scalar metric perturbations can be written in terms of the gauge
invariant metric potentials~\cite{Bardeen80,MFB92}
\begin{eqnarray}
\label{APhiPsi}
\widetilde{A} & \equiv & \widetilde\Phi + \widetilde\Psi
 + \left( {\widetilde\Psi \over \widetilde{h}} \right)' \, ,\\
\label{BPhiPsi}
\widetilde{B} & \equiv & - \, {\widetilde\Psi \over \widetilde{h}} \, .
\end{eqnarray}
Note that the gauge transformation
\begin{equation}
\label{gt}
\eta \to \eta-{\widetilde{\Psi}\over\widetilde{h}} \, ,
\end{equation}
brings the metric of Eq.~(\ref{dds}) into the more commonly used
longitudinal gauge~\cite{MFB92} where
\begin{eqnarray}
d\widetilde{s}^2 & \to & \widetilde{a}^2(\eta)
 \left\{ -(1+2\widetilde{\Phi})d\eta^2 
 + \left[(1-2\widetilde{\Psi})\delta_{ij} + h_{ij}\right]
 dx^i dx^j \right\} \, .
\end{eqnarray}

The curvature perturbation on uniform energy density hypersurfaces (as
$k\eta\to0$) is commonly denoted by $\zeta$~\cite{MFB92} and is given by
\begin{equation}
\zeta \equiv \widetilde{\Phi}
 - {\widetilde{h}^2 \over \widetilde{h}'-\widetilde{h}^2}
 \left( \widetilde\Phi + \widetilde{h}^{-1}\widetilde{\Phi}' \right) 
\,,
\end{equation}
and hence with $\widetilde{h}$ given by Eq.~(\ref{Esf}) for the scale
factor in the Einstein frame, we have
\begin{equation}
\zeta = {\widetilde{A}\over3} \, ,
\end{equation}
in any dilaton-vacuum or axion-dilaton cosmology.

$\zeta$ is a particularly useful quantity to calculate as it becomes
constant on scales much larger than the Hubble scale ($|k\eta|\ll1$) for
purely adiabatic perturbations, even through changes in the equation of
state.  In single-field inflation models this allows one to compute the
density perturbation at late times, during the matter or radiation
dominated eras, by equating $\zeta$ at ``re-entry''
($k=\widetilde{a}\widetilde{H}$) with that at horizon crossing during
inflation.  Thus previous studies have calculated the spectrum of
$\widetilde{A}$, and hence $\zeta$, in order to predict the density
perturbations induced in the pre-Big Bang
scenario~\cite{BrusteinET1995,Hwang}. However, the situation is not
really so straightforward in the pre-Big Bang scenario as in
single-field inflation, because in the full low-energy string effective
action there will be many fields present which can lead to non-adiabatic
perturbations. We must be aware of the fact that density perturbations
at late times may not be simply related to $\zeta$ alone, but may also
be dependent upon fluctuations in other fields. One such field is the
axion field, and we shall see that it may have a markedly different
spectrum from $\zeta$. 

The scalar field perturbations themselves transform under the gauge
transformation $\eta\to\eta+\delta\eta$ giving
$\delta x\to\delta x-x'\delta\eta$. Thus the scalar field
perturbations in the longitudinal gauge ($\delta x_l$) are related to
those in the spatially flat gauge ($\delta x$) under the gauge
transformation in Eq.~(\ref{gt}) as
\begin{equation}
\label{gaugephi}
\delta x \to \delta x_l
 = \delta x + x' {\widetilde\Psi \over \widetilde{h}} \, .
\end{equation}

\subsection{Tensor metric perturbations}

Fortunately, the gravitational wave perturbations $h_{ij}$ are both
gauge and conformally invariant. They decouple from the scalar 
perturbations in the Einstein frame to give a simple evolution equation
for each Fourier mode
\begin{equation}
\label{heom}
h_k'' + 2\widetilde{h}\, h_k' + k^2 h_k  = 0 \, .
\end{equation}
The growing mode in the long wavelength ($|k\eta|\to0$) limit is $h_k
\sim \ln|k\eta|$. (We have not considered gravitational waves
propagating in the $n$ internal dimensions. See
Ref.~\cite{Giovannini}.)  The spectrum depends solely on the dynamics
of the scale factor in the Einstein frame given in Eq.~(\ref{Esf}),
which as we have seen is the same regardless of the time-dependence of
the moduli or axion fields. It leads to a spectrum of primordial
gravitational waves steeply growing on short scales, with a spectral
index $n_T=3$~\cite{pBB,BrusteinET1995}.
This is in contrast to conventional inflation models which require
$n_T<0$~\cite{LiddleLyth}. The graviton spectrum appears to be a
robust and distinctive prediction of any pre-Big Bang type evolution
based upon the low-order string effective action. This has been
discussed extensively elsewhere~\cite{pBB,BrusteinET1995,GW},
so we now turn to discuss in more detail the spectra corresponding to
scalar perturbations. 

\section{Pre-Big Bang spectra}

While the solutions for the homogeneous dilaton, axion and scale
factor in the different frames may lead to interesting behaviour in
the early universe, the success of the standard big bang model
suggests that the evolution should closely approach the
conventional general relativistic evolution at least by the time of
nucleosynthesis. If we are to see any trace of the earlier evolution
it will be in the primordial spectrum of inhomogeneities present on
large-scales that we observe today. Such large-scale structure can
only be generated by some unconventional physics, such as inflation,
topological defects or a pre-Big Bang epoch. During a period of
accelerated expansion the comoving Hubble length $|a/a'|$ decreases
and vacuum fluctuations which are assumed to start in the
flat-spacetime vacuum state may be stretched up to exponentially large
scales. The precise form of the spectrum depends on the expansion of
the homogeneous background and the couplings between the fields.

We have seen that the comoving Hubble length does indeed decrease in
the Einstein frame during the contracting phase when $\eta<0$. Because
the dilaton, moduli fields and graviton are minimally coupled to this
metric, this ensures that small-scale vacuum fluctuations will
eventually be stretched beyond the comoving Hubble scale during this
epoch. 

The production of scalar and tensor metric perturbations in the
pre-Big Bang scenario has been studied by various authors (see for
example~\cite{BrusteinET1995,Hwang}). As we remarked
earlier, the axion field is taken to be a constant in these
solutions. However, while a constant
axion field may be a consistent particular solution when describing
the background classical field, one cannot necessarily neglect quantum
fluctuations in this field. In this section we will consider for the
first time the production of axions during a pre-Big Bang type
evolution (where the background axion field is constant) and then go
on to discuss the perturbation spectrum in the more general case with
$\sigma'\neq0$. We will also analyse the behaviour of these
cosmological vacuum states to first-order under S-duality transformations.

First of all, let us consider the perturbation spectra
produced when the background axion field remains constant,
$\sigma'=0$. The evolution of the homogeneous background fields is
given in Eqs.~(\ref{dilphi}--\ref{dilbeta}) and the dilaton and
moduli fields both evolve as minimally coupled massless fields
in the Einstein frame. In particular, the dilaton perturbations are
decoupled from the axion perturbations and the equations of motion in
the spatially flat gauge, Eq.~(\ref{dphieom}--\ref{dbetaeom}), become
\begin{eqnarray}
\label{pBBdphieom}
\delta\phi'' + 2\widetilde{h}\delta\phi' + k^2\delta\phi
 & = & 0 \, , \\
\label{pBBdsigmaeom}
\delta\sigma''+2\widetilde{h}\delta\sigma' + k^2\delta\sigma
 & = & - 2\phi'\delta\sigma' \\
\label{pBBdbetaeom}
\delta\beta'' +2\widetilde{h}\delta\beta' +k^2\delta\beta
 & = & 0 \, ,
\end{eqnarray}
plus we have the constraint Eq.~(\ref{AofB})
\begin{equation}
\label{pBBAcon}
\widetilde{A} = {\phi'\over4\widetilde{h}} \, \delta\phi
  +{n\beta'\over2\widetilde{h}} \, \delta\beta \, .
\end{equation}

\subsection{Dilaton and moduli perturbations}

{}From Eq.~(\ref{pBBAcon}) we see that, to first-order, the metric
perturbation, $\widetilde{A}$, is determined solely by the dilaton and
moduli field perturbations. 
The canonically normalised field perturbations
are~\cite{Mukhanov88,BrusteinET1995,Giovannini} 
\begin{eqnarray}
\label{defu}
u & \equiv & {1\over\sqrt{2}\kappa} \widetilde{a}\delta\phi \, ,\\
w & \equiv & {\sqrt{n}\over\kappa} \widetilde{a}\delta\beta \, ,
\end{eqnarray}
which, from Eqs.~(\ref{pBBdphieom})
and~(\ref{pBBdbetaeom}), obey the wave equations
\begin{eqnarray}
\label{upp}
u'' + \left( k^2 - {\widetilde{a}''\over\widetilde{a}} \right) u = 0 \, ,\\
w'' + \left( k^2 - {\widetilde{a}''\over\widetilde{a}} \right) w = 0 \, .
\end{eqnarray}
After inserting the simple solution for the Einstein frame
scale factor given in Eqs.~(\ref{Esf}) we find that these equations
give the general solutions
\begin{eqnarray}
\label{usol}
u & = &
 |k\eta|^{1/2} \left[ 
  u_+ H_0^{(1)}(|k\eta|) + u_- H_0^{(2)}(|k\eta|) 
 \right] \, ,\\
w & = &
 |k\eta|^{1/2} \left[ 
  w_+ H_0^{(1)}(|k\eta|) + w_- H_0^{(2)}(|k\eta|)
 \right] \, .
\end{eqnarray}

On the $(+)$ branch, i.e., when $\eta<0$, we can normalise modes at
early times, $\eta\to-\infty$, where all the modes are far inside the
Hubble scale, $k\gg|\eta|^{-1}$, and can be assumed to be in
flat-spacetime vacuum\footnote{It is interesting to note that in conventional
inflation we have to assume that this result for a quantum field in a
classical background holds at the Planck scale. Here, however, the
normalisation is done in the zero-curvature limit in the infinite
past.}. Just as in conventional inflation, this produces perturbations
on scales far outside the horizon, $k\ll|\eta|^{-1}$, at late times,
$\eta\to0$.

Conversely, the solution for the $(-)$ branch with $\eta>0$ is dependent
upon the initial state of modes far outside the horizon,
$k\ll|\eta|^{-1}$, at early times where $\eta\to0$. The role of a period
of inflation, or of the pre-Big Bang $(+)$ branch, is precisely to set
up this initial state which otherwise appears as a mysterious initial
condition in the conventional (non-inflationary) Big Bang model.

Allowing only positive frequency modes in the flat-spacetime vacuum
state at early times for the pre-Big Bang $(+)$ branch
requires~\cite{BirrellDavies} that, as $k\eta\to-\infty$,
\begin{equation}
\label{shortwave}
u \to {e^{-ik\eta} \over \sqrt{2k}} \, ,
\end{equation}
and similarly for $w$, giving
\begin{equation}
\label{uplus}
u_+ = w_+ = e^{i\pi/4} {\sqrt\pi\over2\sqrt{k}} \, , \qquad u_-=w_-=0 \, .
\end{equation}

The power spectrum for perturbations is commonly denoted by
\begin{equation}
{\cal P}_{\delta x} \equiv {k^3\over2\pi^2} |\delta x|^2 \, ,
\end{equation}
and thus for modes far outside the horizon ($k\eta\to0$) we have 
\begin{eqnarray}
\label{pBBphi}
{\cal P}_{\delta\phi} & \to & {4\over\pi^3} \kappa^2\widetilde{H}^2
 (-k\eta)^3[\ln(-k\eta)]^2
 \,,\\
\label{pBBbeta}
{\cal P}_{\delta\beta} & \to & {2\over n\pi^3} \kappa^2\widetilde{H}^2
 (-k\eta)^3[\ln(-k\eta)]^2
 \,,
\end{eqnarray}
where $\widetilde{H}\equiv\widetilde{a}'/\widetilde{a}^2$ is the
Hubble rate in the Einstein frame, 
and recall $n$ is the number of compact dimensions.
The amplitude of the perturbations grows towards small
scales, but only becomes large for modes outside the horizon
($|k\eta|<1$) when $\kappa^2\widetilde{H}^2\sim1$, i.e., the Planck
scale in the Einstein frame.
The spectral tilt of the perturbation spectra is given by
\begin{equation}
\label{specindex}
n_x -1 \equiv {d\ln{\cal P}_{\delta x} \over d\ln k}
\end{equation}
which from Eqs.~(\ref{pBBphi}) and~(\ref{pBBbeta}) gives
$n_\phi=n_\beta=4$ (where we neglect the logarithmic dependence). 

We need also to compute the amplitude of the scalar metric
perturbations, to check the validity of our linear perturbation
analysis. Normalising the amplitude of the spectrum for the metric
perturbation $\widetilde{A}$ in Eq.~(\ref{genA}) from the constraint
Eq.~(\ref{Aconstraint}), using Eqs.~(\ref{dila}--\ref{dilbeta}) for the
background fields and Eqs.(\ref{pBBphi}) and~(\ref{pBBbeta}) for their
perturbations, we have
\begin{equation}
\label{Aspectrum}
{\cal P}_{\widetilde{A}} = {3\over\pi^3}
 \kappa^2\widetilde{H}^2 (-k\eta)^3[\ln(-k\eta)]^2
 \,.
\end{equation}
[Remember that we are adding independent random variables. The $3$ comes
from $r_\pm^2+2ns^2=3$.] Note that this spectrum of scalar metric
perturbations in entirely independent of the integration constants that
parameterise the solutions given in Eqs.~(\ref{dila}--\ref{dilbeta}).
The scalar spectrum, just like the spectrum of tensor perturbations is a
robust prediction of any pre-Big Bang scenario where the universe
collapses in the Einstein frame, and becomes dominated by homogeneous
scalar fields. 

Just like the field perturbations, the scalar metric perturbations have
a steep blue spectrum, $n_{\widetilde{A}}=4$, which becomes large on
superhorizon scales $|k\eta|<1$ only near the Planck scale,
$\kappa^2\widetilde{H}^2\sim1$. Note that Bardeen's gauge invariant
perturbations $\widetilde{\Phi}$ and $\widetilde{\Psi}$, defined in
Eqs.~(\ref{APhiPsi}) and~(\ref{BPhiPsi}), actually become
large much earlier~\cite{BrusteinET1995}, but the fact that the
perturbations remain small in our choice of gauge implies that our
linear calculation is in fact valid up until the
Planck epoch~\cite{BrusteinET1995}.

Unfortunately this leaves us with such a steeply tilted spectrum of
metric perturbations that there would be effectively no primordial
metric perturbations on large (super-galactic) scales in our present
universe if the post-Big Bang era began close to the Planck scale. The
metric fluctuations are of order unity on the Planck scale
($10^{-33}$cm) when $T\sim10^{32}$K in the standard post-Big Bang
model.  This corresponds to a comoving scale of about $0.1$cm today
(when $T=2.7$K), about $10^{-29}$ times the scale of perturbations
observed on the microwave background sky. Thus the microwave
background temperature anisotropies should be of order $10^{-87}$
rather than the observed $10^{-5}$. 
However, it turns out that the presence of the axion field could
provide an alternative spectrum of perturbations more suitable as a
source of large-scale structure.

\subsection{Axion perturbations}

While the dilaton and moduli fields evolve as massless minimally coupled
fields in the Einstein frame, the axion evolves as a massless minimally
coupled field in the axion frame and the canonically normalised field
perturbation is 
\begin{equation}
\label{defv}
v \equiv {1\over\sqrt{2}\kappa} \bar{a}\delta\sigma \, ,
\end{equation}

In this section we are considering the axion spectrum in the pre-Big
Bang scenario where the background axion field is constant. As a result
density perturbations are only second-order in the axion perturbation
and so we can neglect the back-reaction from the metric to linear order.
The field perturbation $\delta\sigma$ is gauge invariant when
$\sigma'=0$ [see Eq.~(\ref{gaugephi})]
and in any gauge, the axion perturbation obeys the
decoupled wave equation given in Eq.~(\ref{pBBdsigmaeom}) which can be
re-written in terms of $v$ as
\begin{equation}
\label{vpp}
v'' + \left( k^2 - {\bar{a}''\over\bar{a}} \right) v = 0 \, .
\end{equation}
As we have just mentioned,
whereas the dilaton and moduli evolve as massless minimally
coupled fields in the Einstein frame, the axion is minimally coupled
in the axion frame, whose evolution given in Eq.~(\ref{axionsf}) is
significantly different. In fact, substituting Eq.~(\ref{axionsf})
in Eq.~(\ref{vpp}) we have
\begin{equation}
\label{vsol}
v  = 
 |k\eta|^{1/2} \left[ 
  v_+ H_{r}^{(1)}(|k\eta|) + v_- H_{r}^{(2)}(|k\eta|) 
 \right] \, ,
\end{equation}
where we have used $r\equiv|r_\pm|$.
Once again, we can only normalise this using the flat spacetime vacuum
state at early times as $-k\eta\to\infty$ on the $(+)$ branch, as in
Eq.~(\ref{shortwave}), which gives
\begin{equation}
\label{vplus}
v_+= e^{i(2r+1)\pi/4} {\sqrt{\pi} \over 2\sqrt{k}} \, , \qquad v_-=0 \, .
\end{equation}
and hence we have
\begin{equation}
\label{pBBdsigma}
\delta\sigma = \kappa \sqrt{{\pi \over 2k}} e^{i(2r+1)\pi/4}
{\sqrt{-k\eta} \over \bar{a}} H_{r}^{(1)}(-k\eta) \,
.
\end{equation}
At late times, as $-k\eta\to0$, we find~\footnote
{When $2ns^2=3$ and $r=0$ the dilaton remains constant and the axion
frame and Einstein frame coincide, upto a constant factor. Thus the
axion spectrum behaves like that for the dilaton and moduli fields and
the late time evolution in this case is that logarithmic with respect
to $-k\eta$, as given in Eqs.~(\ref{pBBphi}) and~(\ref{pBBbeta}).}
\begin{equation}
\label{pBBsigma}
{\cal P}_{\delta\sigma} = 2\kappa^2 \left( {C(r) \over 2\pi} \right)^2
 {k^2\over\bar{a}^2} (-k\eta)^{1-2r} \, ,
\end{equation}
where the numerical coefficient
\begin{equation}
C(r) \equiv {2^r\Gamma(r) \over 2^{3/2}\Gamma(3/2)} \, ,
\end{equation}
approaches unity for $r=3/2$.

The expression for the axion power spectrum can be written in terms of
the field perturbation when each mode crosses outside the horizon
\begin{equation}
{\cal P}_{\delta\sigma_c}
 = 2\kappa^2 \left[{C(r)\over r_\pm+(1/2)}\right]^2
 \left( {\bar{H}_c \over 2\pi} \right)^2 \, ,
\end{equation}
where $\bar{H}_c$ is the Hubble rate when $|k\eta|=1$. 
This is the power spectrum for a
massless scalar field during power-law inflation which approaches the
famous result
${\cal P}_{\delta\sigma}/2\kappa^2=(\bar{H}_c/2\pi)^2$ as $r_-\to-3/2$, and
the expansion in the axion frame becomes exponential\footnote
{The factor $2\kappa^2$ arises due to our dimensionless definition of
$\sigma$.}. 

More importantly, the spectral index
\begin{equation}
n_\sigma = 4 - 2r = 4 - 2\sqrt{3-2ns^2}
\end{equation}
depends crucially upon the value of $r=|r_\pm|$. The spectrum becomes the
classic scale-invariant Harrison-Zel'dovich spectrum as
$r\to3/2$. The lowest possible value of the spectral tilt
$n_\sigma$ is $4-2\sqrt{3}\simeq0.54$ which is obtained when stable
compactification has occurred and the moduli field $\beta$ is fixed. The
more rapidly the internal dimensions evolve, the steeper the resulting
axion spectrum until for $2ns^2=3$ and $r=0$ we have $n_\sigma=4$ like
the dilaton and moduli spectra. Note that the condition for a negatively
tilted spectrum coincides exactly with the requirement for conventional
power-law inflation, rather than pole inflation, in the axion frame.

Of course, when the background axion field is constant these
perturbations, unlike the dilaton or moduli perturbations, do not affect
the scalar metric perturbations (i.e., these are isocurvature
perturbations).  However, if the axion field does affect the energy
density at late times (for instance, by the axion field acquiring a
mass) then the spectrum of density perturbations need not have a steeply
tilted blue spectrum like the dilaton perturbations, but rather could have
a nearly scale-invariant spectrum as required for large-scale
structure formation~\cite{LiddleLyth}.

\section{Perturbation spectra in general axion-dilaton cosmologies}

When we allow the background homogeneous axion field to be
time-dependent we must allow for the interaction between the dilaton
and axion field and the metric to first-order.

In fact we have seen that in the spatially flat gauge the evolution
equations for both the scalar and tensor metric perturbations
[Eqs.~(\ref{Aeom}) and~(\ref{heom})] are independent of the evolution
of the different scalar fields and are determined solely by the
evolution of the Einstein scale factor given in
Eq.~(\ref{Esf}). Because the moduli field perturbations remain
decoupled from both the axion and dilaton, their evolution equation,
Eq.~(\ref{pBBdbetaeom}), is also unaffected. Thus the spectral tilts
of the scalar and tensor metric perturbations and the moduli spectrum,
Eq.~(\ref{pBBbeta}), remain the same as in the pre-Big Bang scenario.

We can understand this in terms of the S-duality transformations that
relate the general axion-dilaton solutions to the dilaton-vacuum
solutions. These transformations leave the Einstein frame metric and
moduli field invariant and thus not only the homogeneous fields, but
also their perturbations, are identical in S-duality related
cosmologies. However the dilaton and axion fields and their
perturbations will in general be affected by S-duality
transformations.

\subsection{Axion and dilaton perturbations}

The dilaton and axion perturbation field equations~(\ref{dphieom})
and~(\ref{dsigmaeom}) become coupled to first order when $\sigma'\neq0$,
and the chances of obtaining analytic solutions might appear to be
remote. However, we can exploit the S-duality symmetry which relates the
general axion-dilaton cosmologies to the much simpler dilaton-vacuum
cosmologies in order to find linear combinations of the axion and
dilaton perturbations which remain straightforward to integrate even in
the more general case.

We define two new S-duality invariant variables
\begin{eqnarray}
\label{defx}
x &\equiv& e^\phi \left( {\phi'\over\widetilde{h}} \delta\sigma
 - {\sigma'\over\widetilde{h}} \delta\phi \right) \,,\\
\label{defy}
y &\equiv& {\phi'\over\widetilde{h}} \delta\phi
 + {e^{2\phi}\sigma'\over\widetilde{h}} \delta\sigma \, .
\end{eqnarray}
In terms of $x$ and $y$ the perturbation equations decouple and the field
equations~(\ref{dphieom}) and~(\ref{dsigmaeom}) then become
\begin{eqnarray}
\label{xeom}
x'' + 2\widetilde{h} x'
 + \left[ k^2 - (\phi'^2+e^{2\phi}\sigma'^2) \right]
 x & = & 0 \\
\label{yeom}
y'' + 2\widetilde{h} y' + k^2 y & = & 0 \, .
\end{eqnarray}

It is far from obvious on first inspection that these variables should 
be invariant under the S-duality transformation given in
Eq.~(\ref{Sduality}).  However written in terms of the matrix $N$
defined in Eq.~(\ref{N}) we have
\begin{eqnarray}
2 \widetilde{h} x & = & {\rm tr}(-JNJN'J\delta N) \,,\\
2 \widetilde{h} y & = & {\rm tr}(JN'J\delta N) \,,
\end{eqnarray}
and we can see that these variables are the unique S-duality invariant 
linear combinations of the axion and dilaton perturbations. They reduce 
to the (decoupled) axion and dilaton perturbations in the pure dilaton-vacuum
background, as $\sigma'\to0$, where we have
\begin{eqnarray}
x & \to & {\phi' \over \widetilde{h}} e^\phi\delta\sigma
=2r_\pm e^\phi\delta\sigma \, ,\\
y & \to & {\phi' \over \widetilde{h}} \delta\phi =2r_\pm \delta\phi \,.
\end{eqnarray}
Note that $x$ is not only S-duality invariant, but also gauge
invariant. That is, it doesn't matter which gauge we choose to
calculate $\delta\sigma$ and $\delta\phi$, the combination which
defines $x$ remains unchanged. It is proportional to the axion
perturbation on uniform-dilaton hypersurfaces,
$\delta\sigma|_\phi\equiv \delta\sigma -\sigma'(\delta\phi/\phi')$. By
symmetry, it is also proportional to the dilaton perturbation on
constant axion hypersurfaces, though this perturbation diverges in the
limit that the background axion field becomes constant.

Having found S-duality invariant variables, one can verify that the
evolution equations for these variables, Eqs.~(\ref{xeom})
and~(\ref{yeom}) are themselves invariant under S-duality. Remembering
that the general axion-dilaton cosmological solutions can always be
related to the dilaton-vacuum solutions by an S-duality transform, we
see that the evolution equations for $x$ and $y$ in an arbitrary
axion-dilaton cosmology are exactly the same as those for the axion and
dilaton perturbations in the dilaton-vacuum case. Just as in the
constant axion case, we can define canonically normalised variables
\begin{eqnarray}
u &\equiv& {1\over2r\sqrt{2}\kappa} \widetilde{a} y \, ,\\
v &\equiv& {1\over2r\sqrt{2}\kappa} \widetilde{a} x \, ,
\end{eqnarray}
which coincide with the definitions given in Eqs.~(\ref{defu})
and~(\ref{defv}) in the dilaton-vacuum case.
In general, $u$ obeys the S-duality invariant equation of motion given in
Eq.~(\ref{upp}) and whose general solution is given by
Eq.~(\ref{usol}). The equation of motion for $v$ given in Eq.~(\ref{vpp}),
however, is not invariant under an S-duality transformation. Instead
the S-duality invariant version of the equation of motion is
\begin{equation}
v'' + \left( k^2 - {r^2-1/4 \over \eta^2} \right) v = 0 \, .
\end{equation}
which coincides with Eq.~(\ref{vpp}) when $\sigma'=0$. The general
solution for $v$ is thus still given by Eq.~(\ref{vsol}).

We can still normalise cosmological vacuum perturbations at early
times on the $(+)$ branch as $\eta\to-\infty$ because we have seen
that in this limit the general axion-dilaton solution given in
Eqs.~(\ref{axiphi}--\ref{axisigma}) approach the constant axion
solutions with $r_+=+r$. Thus the constants $u_\pm$ and $v_\pm$ are
given by Eqs.~(\ref{uplus}) and~(\ref{vplus}). 
By picking S-duality invariant field perturbations we have been able to
calculate the general axion-dilaton cosmological perturbation spectra
using the pure dilaton-vacuum cosmological vacuum states.
We have
\begin{equation}
\label{Py}
{\cal P}_y \to {16r^2\over\pi^3} \kappa^2\widetilde{H}^2
 (-k\eta)^3[\ln(-k\eta)]^2
 \,.
\end{equation}
and the generalised axion perturbation spectrum is given by
\begin{equation}
{\cal P}_x \to 8r^2\kappa^2 \left( {C(r) \over 2\pi} \right)^2
 {k^2\over\widetilde{a}^2} (-k\eta)^{1-2r} \, .
\end{equation}

To recover the actual (though gauge and S-duality dependent) axion and
dilaton perturbations we can invert Eqs.~(\ref{defx}) and~(\ref{defy})
to give
\begin{eqnarray}
\delta\sigma & = & {e^{-\phi} \over 4r^2}
 \left( {\phi'\over\widetilde{h}} x +
  {e^\phi\sigma' \over \widetilde{h}} y \right) \,,\\
\delta\phi & = & {1 \over 4r^2}
 \left( {\phi'\over\widetilde{h}} y -
  {e^\phi\sigma' \over \widetilde{h}} x \right) \,.
\end{eqnarray}

However, at late times on the $(+)$ branch, as $\eta\to0$ the general
axion-dilaton solutions approach dilaton-vacuum solutions with $r_-=-r$,
and hence $\delta\phi\to y/2r_-$ and $\delta\sigma\to e^{-\phi}x/2r_-$.
Note that the change of sign from $r_+=+r$ to $r_-=-r$ between the early
and late time dilaton-vacuum solutions leads to a phase shift $e^{i\pi}$
with respect to the late-time behaviour of the pure dilaton-vacuum
solutions. But the final power spectrum for the dilaton and axion
perturbations as $\eta\to0$ in the general axion-dilaton cosmologies is
identical to that given in Eqs.~(\ref{pBBphi}) and~(\ref{pBBsigma}) for
the S-duality related dilaton-vacuum case. The tilt and amplitude of the
spectra are determined solely by the parameter $r=|r_\pm|$ and are
insensitive to the specific time dependence of the axion field in
different, but S-duality related, solutions.

The constraint equation for $\widetilde{A}$,
Eq.~(\ref{Aconstraint}), includes only $y$ and $\delta\beta$;
\begin{equation}
 \widetilde{A} = {1\over4}y +
 {n\beta'\over2\widetilde{h}}\delta\beta \, .
\end{equation}
{}From Eqs.~(\ref{Py}) and~(\ref{pBBbeta})
we see that the spectrum of scalar metric perturbations 
is unaffected by the time dependence of the axion field and is the same
as that obtained in the constant axion case, given in
Eq.~(\ref{Aspectrum}).  

\section{Discussion}

The low-energy limit of string theory, or M-theory, contains many
different degrees of freedom. In this paper we have considered a very
simple model containing only a $D=4$ spatially flat FRW metric with a
dilaton, a single massless modulus field (representing the volume of $n$
internal dimensions) and a pseudo-scalar axion field derived from the
Neveu-Schwarz antisymmetric tensor potential. The axion-dilaton
solutions can be generated from the dilaton-vacuum solutions by an
S-duality transformation. They 
generalise the power-law dilaton-vacuum solutions in a particularly
simple way, interpolating between two asymptotically dilaton-vacuum
regimes, which are themselves related by an S-duality transformation.

Although the general axion-dilaton solutions do not alter the singular
nature of the cosmological solutions in the string or Einstein frame, we
draw attention to the fact that the evolution in the conformally related
axion frame (in which the axion is minimally coupled) can become
non-singular when the axion field is allowed to be 
time-dependent. The worldlines of axionic observers can have an infinite
proper lifetime in this frame. There is no graceful
exit~\cite{Bru+Ven94} from the pre-Big Bang $(+)$ branch to the post-Big
Bang $(-)$ branch, but the $(+)$ or $(-)$ branches
themselves can have an infinite proper lifetime.

The shrinking comoving Hubble length during a pre-Big Bang era
generates a spectrum of perturbations about the homogeneous background
fields from quantum fluctuations. 
We have calculated the spectrum of large scale
perturbations produced in the axion field. 
The axion spectral index can lie anywhere in
the range $0.54$ to $4$, which includes the possibility of the nearly
scale-invariant ($n\sim1$) spectrum required for structure formation.
This is in contrast to the dilaton and moduli perturbations 
which have a steep blue spectrum with an index of $n=4$, making them
incapable of seeding large scale structure in our present universe. 
The actual value of the
axion spectral index depends on the rate of expansion of the  internal
dimensions. If stable compactification has already occured,  leading to
an effective $4$-dimensional spacetime, the spectral index  is $n=0.54$.

In the simplest case where the background axion field is
constant, the axion perturbations are isocurvature perturbations during
the pre-Big Bang epoch. Whether these axion perturbations are able to
seed large-scale  structure in the post-Big Bang universe depends
crucially on the coupling between the axion and the matter which dominates the
universe today. 
Nonetheless, it is intriguing that, in principle, the axion could give
rise to a nearly scale-invariant spectrum and that the tilt of that
spectrum is dependent on the compactification of the internal
dimensions. 

We have seen that S-duality is a powerful tool for calculating not
only the classical background solutions in general axion-dilaton
cosmologies but  also the semi-classical perturbation spectra. By
constructing explicitly S-duality invariant field perturbations we
are able to calculate the perturbation spectra in the more
general axion-dilaton cosmologies as well as the dilaton-vacuum case. It
is not surprising that by taking S-duality invariant field perturbations
we can derive S-duality invariant solutions.  More remarkably, however,
the late time dilaton and axion spectra  turn out to be independent of
the preceding evolution along different,  but S-duality related, classical
solutions. 
This results from the fact that S-duality related axion-dilaton
solutions all approach the same dilaton-vacuum solution at late times.
By contrast, other symmetries present in the low energy string action, such
as the symmetry which mixes the moduli field with the dilaton and
axion~\cite{Jim}, do not relate solutions with the same late
time behaviour and so will not leave the perturbation spectra
invariant.

Despite the well-known
problems associated with achieving a graceful exit from the pre-Big
Bang era, it is worth noting a couple of advantages that the pre-Big Bang
predictions have over conventional (potential-dominated) inflation
models. Firstly, the perturbations originate as vacuum fluctuations at
early times, where their amplitude is normalised in a low-curvature,
weakly coupled regime in the infinite past, and not at arbitrarily small
scales during the Planck epoch when the correct vacuum state may be
uncertain.  Secondly, one can give analytic expressions for the
asymptotic perturbations on large scales without having to invoke any
slow-roll type approximations as must usually be done in conventional
inflation models. This is possible not only in the presence of the
dilaton alone, but also when one incorporates the moduli and axion
fields.

\section{Acknowledgements}
We are grateful to B. de Carlos, M. Giovannini, R. Gregory, and J. Lidsey
for useful conversations. DW and RE thank Sussex, and RE thanks Portsmouth, 
Cornell and Waikato universities for their hospitality
while this work was in progress. RE was supported by the Japan
Society for the Promotion of Science. EJC is grateful to the 
physics department at Dartmouth College for their hospitality 
while some of this work was carried out.

\newpage
\section*{Figure Captions}

\vskip12pt
\noindent
{\bf Figure 1:} The string frame scale factor for the axion-dilaton
solution (solid line) given in Eq.~(\ref{axia}). It approaches the
$r_-=-r$ dilaton-vacuum solution given in Eq.~(\ref{dila}) (dashed line)
as $\eta\rightarrow 0$ and the $r_+=+r$ dilaton-vacuum solution
(dot-dashed line) as $\eta \rightarrow \pm\infty$. The $(+)$ and $(-)$
branches are also labelled. The specific parameters for the
axion-dilaton solution are $a_* = 1$, $r=1.2$ and $\eta_* = 2$.

\vskip12pt
\noindent
{\bf Figure 2:} The dilaton, $\phi$, with $\phi_* = 0$ and
the same notation and parameter values as in Fig.~1. 
The axion-dilaton solution [Eq.~(\ref{axiphi})] approaches the $r_-$
dilaton-vacuum solution [Eq.~(\ref{dilphi})] at small $|\eta|$ and the
$r_+$ solution at large $|\eta|$.

\vskip12pt
\noindent
{\bf Figure 3:} The modulus, $\beta$, with $n=6$ and the same
parameter values as in Fig.~1 and $s = \sqrt{13}/10$. The evolution
is the same, for both the dilaton-vacuum in Eq.~(\ref{dilbeta}) and
axion-dilaton solutions in Eq.~(\ref{axibeta}).

\vskip12pt
\noindent
{\bf Figure 4:} The axion, $\sigma$, of the axion-dilaton solution given in
Eq.~(\ref{axisigma}), with the same parameter values as in Fig.~1 and
$\sigma_* = 1$.

\vskip12pt
\noindent
{\bf Figure 5:} The Einstein frame scale factor, $\widetilde{a}$,
plotted against proper time in the Einstein frame for the
dilaton-vacuum and axion-dilaton solutions displayed in the previous
figures.

\vskip12pt
\noindent
{\bf Figure 6:} The axion frame scale factor, $\bar{a}$, plotted
against proper time in the axion frame for axion-dilaton cosmologies
(solid lines) with three different values of $r$, while the other
constants are the same as in Fig.~1. The asymptotic
dilaton-vacuum solutions with $r_-=-r$ (dashed lines) and $r_+=+r$
(dot-dashed lines) are also shown. When $r\geq1.5$ the semi-infinite
interval in $\eta$ is mapped onto an unbounded interval for the proper
time, so the $(+)$ and $(-)$ branch solutions are displayed separately
in the two top figures.

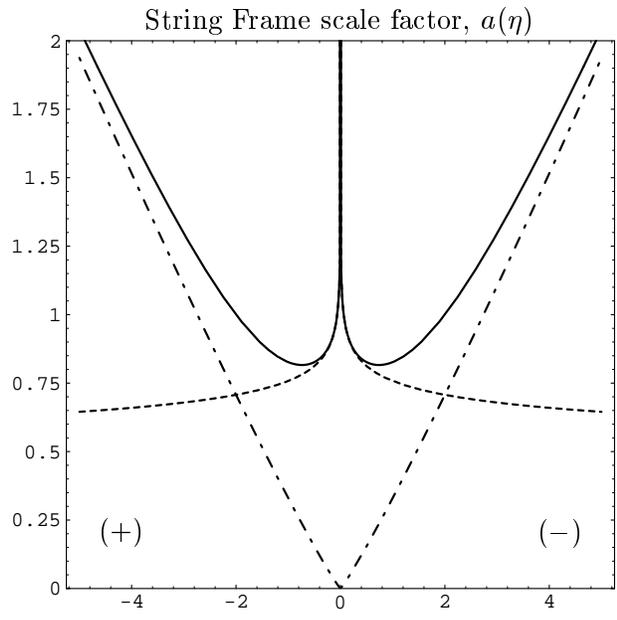
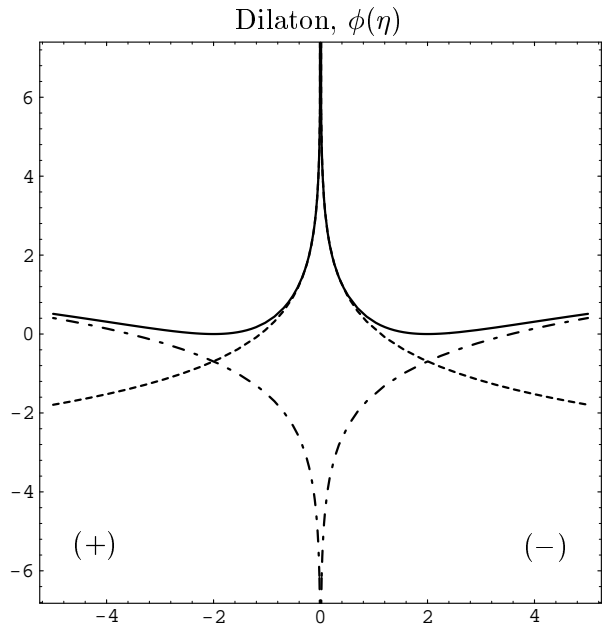
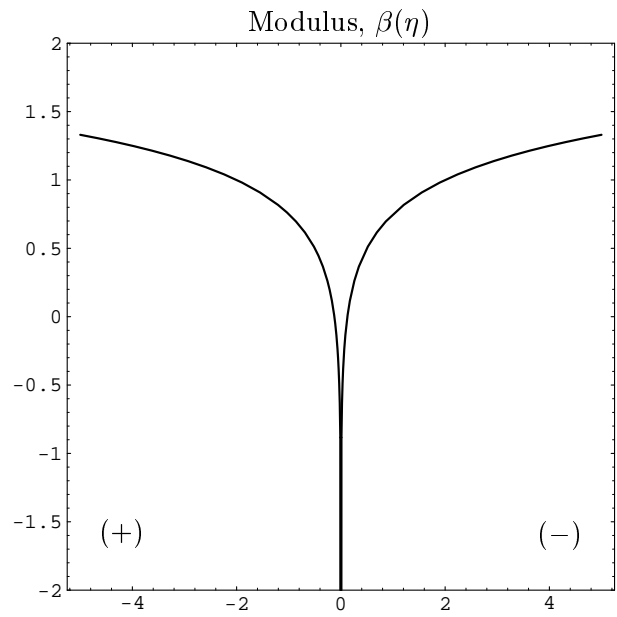
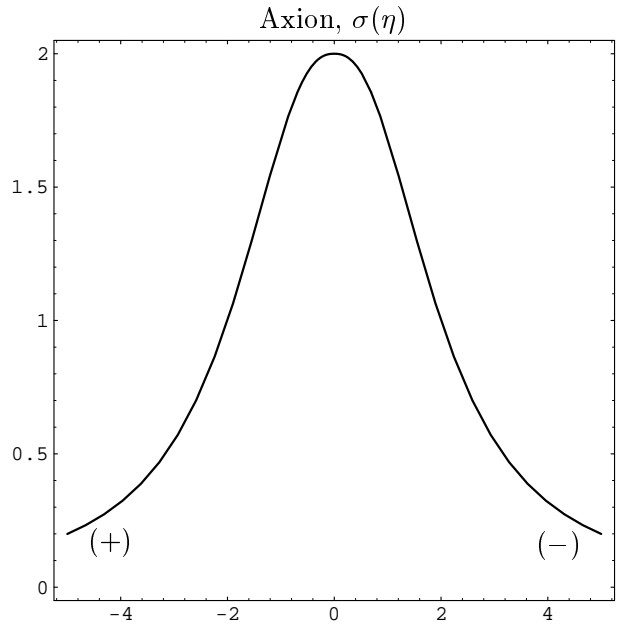
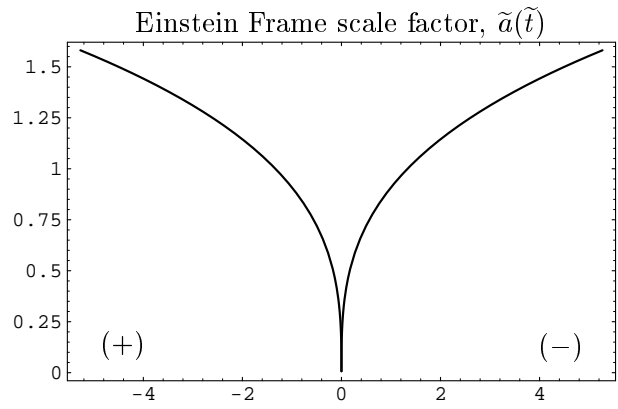
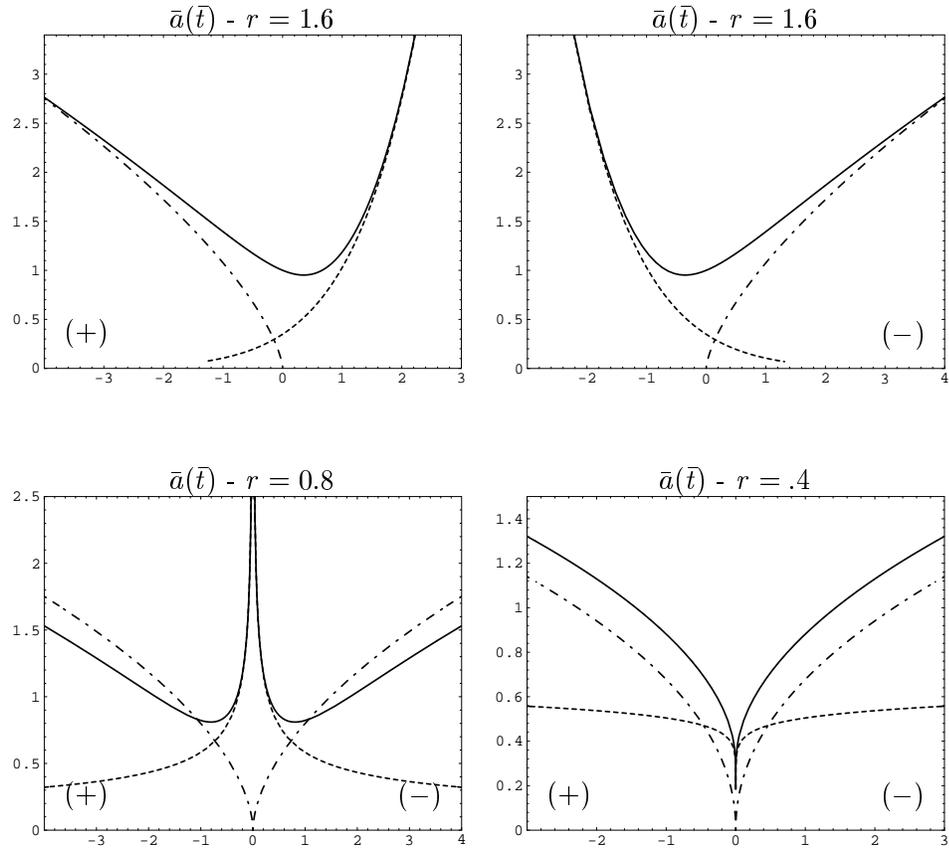
\end{document}